\documentclass{iopjournal}

\usepackage{amsmath}
\usepackage{amssymb}
\usepackage{graphicx}
\usepackage{dcolumn}
\usepackage{bm}
\usepackage{mathtools}
\usepackage{xcolor}
\usepackage{hyperref}
\usepackage[sort]{cite}

\renewcommand{\vec}[1]{{\bf #1}}
\newcommand{\sub}[1]{_{\mbox{\scriptsize #1}}}
\DeclareMathOperator{\cov}{Cov}
\DeclareMathOperator{\var}{Var}

\begin{document}

\articletype{Paper} 

\title{Practical Ellipse Fitting for Atom Interferometry}

\author{N. Fowler$^{1,\dagger}$\orcid{0009-0007-9239-5933}, D. B. Reinhardt$^{2,\dagger}$\orcid{0009-0004-0981-2838}, M. Meister$^2$\orcid{0000-0001-7210-8588} and C. A. Sackett$^{1,*}$\orcid{0000-0001-6741-0571}}

\affil{$^1$Department of Physics, University of Virginia, Charlottesville, Virginia 22904, USA}

\affil{$^2$German Aerospace Center (DLR), Institute of Quantum Technologies, 89081, Ulm, Germany}

\affil{$^*$Author to whom any correspondence should be addressed.}
\affil{$^\dagger$These authors contributed equally to this work.}

\email{sackett@virginia.edu}


\begin{abstract}
In differential atom interferometry, it is necessary to extract the differential phase shift
between two interferometers that are individually noisy. A variety of techniques can be
used for this purpose, under the general label of ellipse fitting. We compare several
methods for ellipse fitting, considering parameters and performance relevant to 
atom interferometry. Statistical, algebraic, geometric and Bayesian methods are discussed,
and we describe a novel 'reciprocal' geometric method that offers some benefits. 
Each method has advantages and disadvantages that depend on the use case required. 
Supplemental material includes basic Python and Matlab code for each method.
\end{abstract}

\section{Introduction}

Atom interferometry \cite{berman1997atom,cronin2007atom} is a powerful measurement technique with applications ranging
from solving practical problems like inertial navigation \cite{d2024atom} to probing questions in fundamental physics like
the nature of dark matter \cite{buchmueller2023large}.
An atom interferometer works by quantum mechanically splitting an atomic wave function into two spatially separated arms that both interact differently
with the environment. When the arms are recombined, the resulting population distribution 
depends on the quantum phase that has developed between them. 
For applications, the physical quantity of interest is encoded into this interferometer phase. 

In some cases, however, the measured phase of an interferometer is noisy due to uncontrolled environmental effects \cite{Bongs2019,meister2026magnetometry}.
For instance, a measurement of gravity on a moving vehicle will suffer from noise due to vibrational
accelerations \cite{Geiger2011, Becker2018, Bidel2018, Narducci2022, Zhang2023}. 
If the phase variation from the noise exceeds $2\pi$ radians, then no information
about gravity can be obtained since the output of the interferometer will be effectively random. 
One way to circumvent this problem is to implement two simultaneous interferometers that are both subject to the
same noise, but with different sensitivity to the quantity of interest. In the case of gravimetry, two interferometers at different
vertical locations can be used to measure the gravity gradient even in the presence of noise \cite{snadden1998measurement}.
Other applications of this differential interferometry
technique include the use of counter-propagating Sagnac interferometers for rotation sensing \cite{Gustavson1997} and the use of 
different atomic species for tests of the Einstein equivalence principle \cite{altschul2015quantum, bonnin2015characterization, asenbaum2020atom}. 

In order to use the differential interferometry approach, it is necessary to extract the nominally constant
differential phase from the correlations between the two noisy interferometer outputs. A conventional method
to achieve this is to plot the set of output values against each other in a graph, where the points will form
an ellipse whose ellipticity depends on the differential phase. By fitting the points to an ellipse, the
phase can be determined. However, fitting points to an ellipse is less straightforward than might be expected.

Ellipse fitting techniques are used in a variety of fields, including
image processing and computer vision \cite{kanatani2016ellipse},
optical signal analysis \cite{heydemann1981determination,povzar2011enhanced, Watkins2014Part2,qin2024homodyne},
and characterization of strain in materials \cite{mulchrone2004fitting}.
A wide variety of techniques have been developed, none of which are flawless \cite{wynn2005comparative,Collett2014Part1,kanatani2016ellipse}.
It can be challenging to determine which technique is optimal in a given situation.

In comparison to other applications, atom interferometry data can exhibit relatively high noise levels and
can have slow acquisition rates, leading to relatively small data sets. For this reason, 
rapid analysis speed can be less important.
Also, interferometry applications
are primarily concerned with determination of the differential phase, in comparison to 
other parameters like the fringe offsets and amplitudes, or to geometric
quantities like the ellipse axes.
Some prior works have considered ellipse fitting specifically in the context of atom interferometry \cite{Foster2002,stockton2007bayesian,sorrentino2010sensitive,Zhang2023,Ridley2024}, 
but have focused on a limited set of methods and noise sources.

This paper aims to provide a practical overview and comparison of several analysis techniques, including statistical \cite{Reinhardt2026prep}, 
algebraic \cite{fitzgibbon1999direct}, geometric \cite{Gander1994}, and Bayesian \cite{stockton2007bayesian}. 
We focus primarily on two critical factors: First, the extracted differential phase
is in general biased away from the true value, limiting the accuracy achievable in such a measurement. 
This impacts the accuracy of the interferometric measurement.
Second, the extracted phase typically fluctuates more
than would be expected for a standard single-interferometer
measurement with the equivalent amount of noise in the signal. This impacts the precision
of the measurement. We consider how both of these factors depend on 
the type of noise and the analysis method. 
All results presented here are computational, based on analysis of synthetic data, but the authors have applied these
methods to experimental data in previous work \cite{beydler2024guided,williams2024pathfinder,meister2026magnetometry}.

As part of this analysis, we present a variation on the geometric fitting technique that we have not 
seen discussed in the literature. It uses a discretization approach that significantly reduces the
required analysis time, and also enables a ``reciprocal'' technique that involves minimizing
both the distance of the data points to the fit curve and the distance from the fit curve to the data points.
This can moderately reduce the phase bias while substantially improving the performance for the offset and
amplitude parameters when these are needed. We also present a method to infer the signal noise level
from the quality of the fit, which offers some advantages over previously-used methods.

The paper is organized by method. Section II describes the data and noise models used, and gives an overview of
the analysis results. Section III describes the statistical and algebraic methods, which are very fast but suffer from relatively large
biases. Section IV describes geometric methods, which are slower but can perform better and offer the most straightforward
physical interpretation. Section V describes Bayesian methods, which can perform very well but typically require other
methods to fix some of the analysis parameters. Finally, Section VI summarizes our results and offers recommendations
for deciding between methods.

\section{Models and Methods}

\subsection{Model Signals}

We consider two interferometer signals labeled $x$ and $y$, which take the general form
\begin{align} \label{model}
x & = c_x + a_x \cos(\xi + \phi/2) \nonumber \\
y & = c_y + a_y \cos(\xi - \phi/2),
\end{align}
with offsets $c_i$, amplitudes $a_i$ and differential phase $\phi$. The common phase $\xi$ is assumed to vary randomly and
uniformly between 0 and $2\pi$. Direct substitution shows that these functions satisfy the equation of an ellipse,
\begin{equation} \label{ellipse}
\frac{(x-c_x)^2}{a_x^2} + \frac{(y-c_y)^2}{a_y^2} - \frac{2(x-c_x)(y-c_y) \cos\phi}{a_x a_y} = \sin^2\phi,
\end{equation}
with center $(c_x, c_y)$, axes $\left[a_x^2 + a_y^2 \pm (a_x^4\! +\! a_y^4\! +\! 2a_x^2 a_y^2 \cos 2\phi)^{1/2}\right]^{1/2}\!\!/2^{1/2}$, and
major axis angle $(1/2) \tan^{-1}\left[2a_x a_y \cos\phi/(a_x^2-a_y^2)\right]$.
Since Eq.~\eqref{ellipse} is invariant
under $\phi \rightarrow 2\pi - \phi$, the phase can only be determined modulo $\pi$ and we take its range to be $0 \leq \phi \leq \pi$. Further, since ellipses with phase $\phi$ and $\pi-\phi$ are identical but for a 
$90^\circ$ rotation, our analysis here will usually be restricted to $0\leq \phi \leq \pi/2$.

The assumption that $\xi$ varies uniformly over its full range is not trivial, and it significantly simplifies
the analysis problem we consider. In practice, it is usually possible to ensure this condition by introducing artificial 
noise if the natural phase variations are not sufficiently large. In such a case, however, it may be more effective
to use an analysis technique tailored to the low-noise data, as opposed to the methods discussed here.

Any of the ellipse parameters can be subject to noise. Given nominal values for the offsets, amplitudes, and differential phase, we construct a synthetic data point by selecting a random value for $\xi$.
Based on the noise model being tested, we then add Gaussian random variables $s$ to one or more
of these parameters, using the probability distribution
\begin{equation}
P(s) = \frac{1}{\sqrt{2\pi\sigma}} e^{-s^2/2\sigma^2}
\end{equation}
where $\sigma$ is the noise deviation. For offset noise, we apply independent noise to $c_x$ and $c_y$:
physically, this can arise from noise in the atom detection system, which is often the
dominant noise source. For phase noise we apply noise to $\phi$: physically this can represent either
quantum projection noise or classical noise in the quantity being measured by the interferometer.
For amplitude noise, we apply independent noise to $a_x$ and $a_y$: physically this can represent
variations in the laser beam alignment or atomic motion that impact the final state overlap
in the interferometer. 
While amplitude noise does occur in practice, it is rarely dominant. Also, we find the impact of 
amplitude noise on phase estimation is generally similar to that of offset noise, while the impact
of phase noise is more distinct.
For these reasons, we focus on offset noise and phase noise here.

To further simplify, we assume nominal values $c_x = c_y = 0$ and $a_x = a_y = 1$.
An experimental data set that does not satisfy these conditions can always be recentered and rescaled
to do so. Recentering has no effect on the model, but rescaling the amplitudes requires
the offset and amplitude noise deviations to be similarly rescaled. We also assume that
the noise deviations for the $x$ and $y$ coordinates are identical, reflecting the fact that the two
interferometers are typically constructed to be as similar as possible. Small deviations from any of these
assumptions have negligible impact on the fitting performance.

\subsection{Analysis Method}

\begin{figure}
	\includegraphics[width=0.6\textwidth]{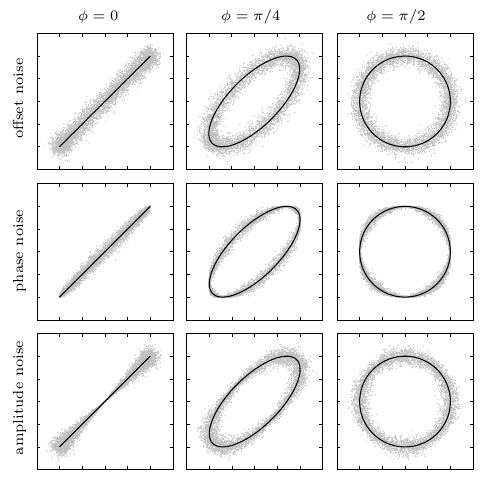}
    \centering
	\caption{\label{fig_examples}
    Example differential interferometer data sets. For each graph, the horizontal shows the $x$ signal and the
    vertical axis shows the $y$ signal. Gray points are synthetic data, and black curves show the underlying model
    with no noise. All sets use $n = 5000$ points with a noise deviation $\sigma = 0.1$. Columns show the labeled 
    differential phase values $\phi$, and rows show the labeled noise models.}
\end{figure}

The analysis procedure used throughout this paper is to generate $K$ synthetic data sets, each
consisting of $n$ data points obtained as described above. Figure~\ref{fig_examples} shows example data sets 
with large $n$ for a few different phase values and noise types.
Each data set is analyzed using a specified method,
generating $K$ sets of parameter estimates $\{c_{x,k}, c_{y,k}, a_{x,k}, a_{y,k}, \phi_k\}$ with $k=1,...,K$. 
Focusing first on the phase, we define the bias 
\begin{equation}
\Delta\phi \equiv \langle \phi_k \rangle - \phi,
\end{equation}
where the average is over all $K$ parameter estimates and $\phi$ is the known model phase value used to create the data set.
We likewise define the phase standard deviation 
\begin{equation}
\delta\phi \equiv \left[\langle \phi_k^2 \rangle - \langle \phi_k \rangle^2\right]^{1/2}.
\end{equation}
We use a similar notation for the offsets' and amplitudes' bias and deviation, 
where we can consider the union of the signals $x$ and $y$ values in the averages since they are assumed symmetric. 
While the extracted phase is of primary interest, the center and amplitude accuracies can also be important,
such as when these parameters are needed for a subsequent analysis step.

As noted, phase noise can be used to represent quantum projection noise. In a 
single interferometer using $N\gg 1$ particles, projection noise is equivalent to phase noise
with $\sigma = 1/\sqrt{N}$ \cite{itano1993quantum}. 
For a differential interferometer using $N$ particles total,
each single interferometer will use half the atoms and therefore have phase noise 
with $\sigma_0 = \sqrt{2/N}$. When the two
phases are subtracted to obtain the differential phase, the independent variances $\sigma_0^2$ will add,
resulting in a differential phase deviation $\sigma = \sqrt{2\sigma_0^2} = 2/\sqrt{N}$. 
Notably, this is a factor of two larger than the normal
standard quantum limit for $N$ atoms.

As will be seen, it is generally not straightforward to achieve this limit even if the
measurements are subject only to projection noise. We do, however, find that the 
parameter deviations $\delta\phi$, $\delta c$ and $\delta a$ all scale directly with
the noise deviation $\sigma$ and inversely with the square root of the number of data points
$n$. To reflect this, we report scaled deviations $\delta\tilde\phi = \delta\phi\cdot \sqrt{n}/\sigma$,
and similarly for $\delta\tilde c$ and $\delta\tilde a$. An analysis method capable of matching
the projection
noise limit would have $\delta\tilde\phi = 1$.

\subsection{Phase Bias}

Unlike in a standard interferometer, the differential measurement is generally subject to bias in
the presence of noise. This is easiest to understand for differential phases near zero or $\pi$,
where in the absence of noise the data points would fall on a straight line. In the presence of
noise, the points instead fall near a line, where they naturally fit to a
non-degenerate ellipse; the phase noise plot in Fig.~\ref{fig_examples} shows a clear example. 
This bias is inherent to any analysis technique: 
consider data with a true phase $\phi = 0$ in the presence of phase noise with
deviation $\sigma$. The actual phase occurring for any given measurement will then range (roughly) from
$-\sigma$ to $+\sigma$. However, since the estimated phase is restricted to be positive, the negative phase
samples will be aliased to positive values, and the recovered phase will instead range from 0 to $+\sigma$.
On average, this results in a bias $\Delta\phi \sim \sigma/2$. 
This bias persists to a lesser degree away from the degenerate points, 
and will generally be zero only at the symmetric point $\phi = \pi/2$. 

\begin{figure}
	\includegraphics[width=0.5\textwidth]{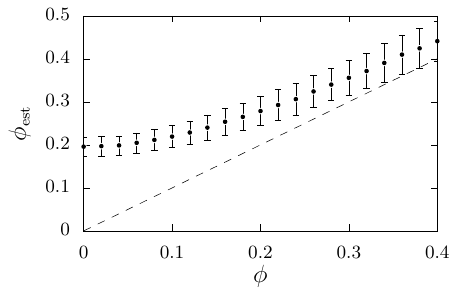}
    \centering
	\caption{\label{phasecurve}
    Estimated phase $\phi\sub{est}$ as a function of the true phase $\phi$ used to generate the synthetic data set.
    The data sets use $n=50$ points and offset noise with $\sigma = 0.1$. Points are averages of the phases obtained from
    $K=10^4$ datasets, and error bars show the standard deviations of the phases, $\delta\phi$.
    The dashed line shows $\phi\sub{est}=\phi$ for reference. 
    The estimated phase is determined using the statistical method of Section \protect\ref{secalg}, 
    but qualitatively similar results are observed for all the methods we considered.
    }
\end{figure}

For this reason, operating at $\phi = \pi/2$ is generally optimal, but doing so is
not always practical. For instance, the gravity gradient may not be large enough to give
a $\pi/2$ phase difference in a gradiometer. In such cases, understanding
and minimizing the bias is critical. We present phase bias calculations for all the methods
considered here, which can in principle be subtracted from extracted phases to improve accuracy. However,
this does not improve sensing performance for measurements at $\phi = 0$ or $\pi$, because
the bias variations are always such that the estimated phase
approaches a constant as the true phase approaches zero.
An example can be seen in Fig.~\ref{phasecurve}. Due to this effect, small variations in 
$\phi$ around $\phi = 0$ cannot be detected efficiently even when the zero-phase bias is known.

A similar bias problem applies to amplitude measurements. In this case the bias is symmetric around
$\phi = \pi/2$ and is thus not necessarily zero anywhere. In practice, however, the bias is typically
smallest at $\pi/2$.
None of the methods considered give a bias for the offset, so we do not report $\Delta c$ values.

\begin{figure*}
	\includegraphics[width=1.\textwidth]{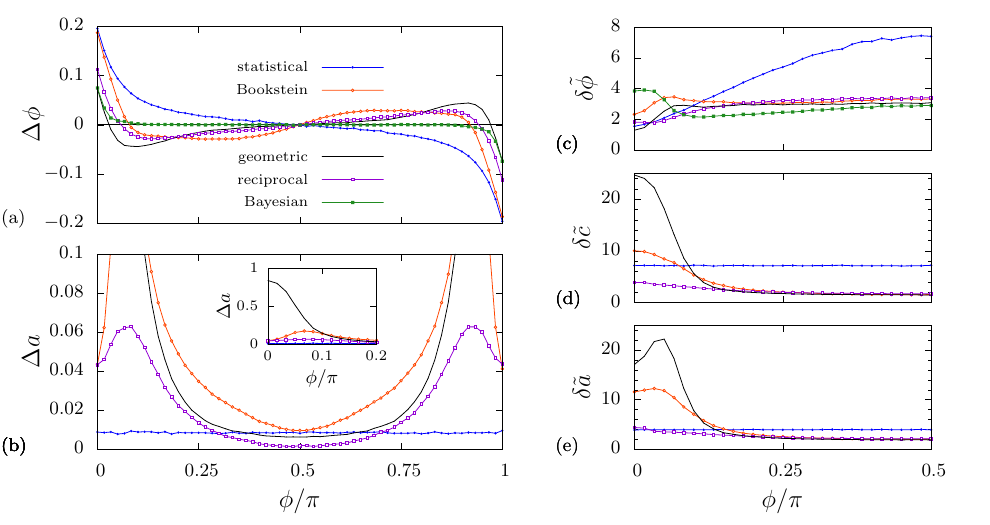}
    \centering
	\caption{\label{fig_offset}
    Comparison of fitting methods, using $n = 50$ data points and offset noise with $\sigma = 0.1$. 
    Calculated values are shown as data points connected by lines for clarity.
    (a) Phase bias $\Delta\phi = \langle \phi \rangle - \phi$ vs.~actual differential phase $\phi$.
    (b) Amplitude bias $\Delta a = \langle a \rangle - a$. The inset shows the behavior at small $\phi$.
    (c) Scaled phase deviation $\delta\tilde\phi = (\langle \phi^2 \rangle - \langle\phi\rangle^2)^{1/2}\cdot \sqrt{n}/\sigma$. 
    (d) Scaled offset deviation $\delta\tilde c = (\langle c^2 \rangle - \langle c\rangle^2)^{1/2}\cdot \sqrt{n}/\sigma$. 
    (e) Scaled amplitude deviation $\delta\tilde a = (\langle a^2 \rangle - \langle a\rangle^2)^{1/2}\cdot \sqrt{n}/\sigma$. 
    The color legend in (a) applies to all plots. Plots (c), (d) and (e) show $0\leq \phi \leq \pi/2$ because the 
    deviations are symmetric around $\phi = \pi/2$, as in (b). Bayesian results are only shown in (a) and (c) because
    our implementation of this method provides only phase values. Averages are over $K = 10^4$ data sets.}
\end{figure*}

\begin{figure*}
	\includegraphics[width=1.\textwidth]{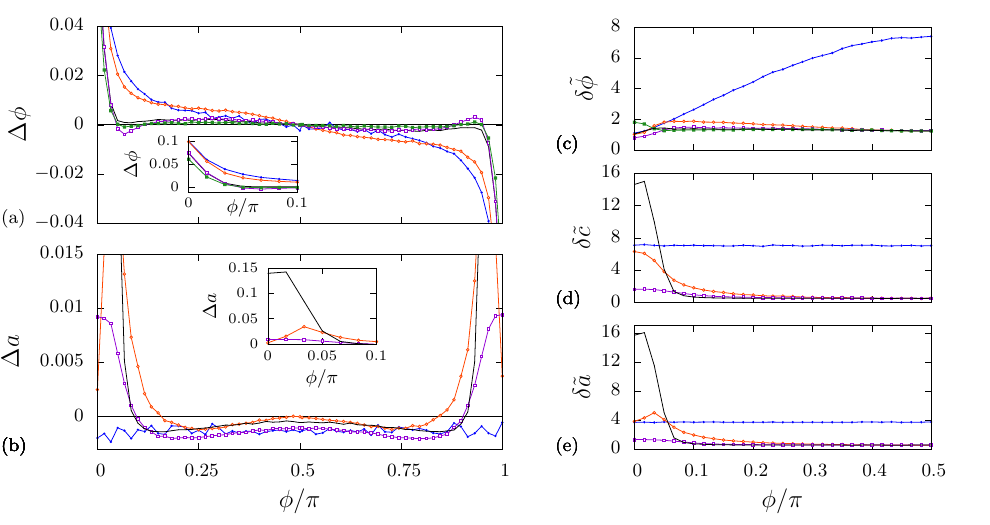}
    \centering
	\caption{\label{fig_phase}
    Comparison of fitting methods in the presence of phase noise. Parameters, colors, and point styles
    are otherwise the same as in Fig.~\protect\ref{fig_offset}.}
\end{figure*}

\subsection{Summary of Results}

Figures~\ref{fig_offset} and \ref{fig_phase} show representative results for the bias and deviation
parameters using five analysis methods that we highlight. Table~\ref{tab_times} gives
the relative computation speeds. The statistical method (blue) is very fast and relies
on statistical analysis of the data rather than any kind of fitting procedure. It 
is relatively noisy but maintains reasonable performance across the full range of phases.
The Bookstein method (red) is an algebraic technique that is also fast and improves the noise
performance. The geometrical method (black) is very slow. It provides good accuracy for
the phase near the degenerate points, but has poor performance for the amplitude and offset 
in that region. The reciprocal method (purple) is our modified geometrical technique that runs faster
and generally performs better. Finally the Bayesian method (green) gives low phase bias values
but has higher noise at near the degenerate points. Also, the Bayesian method we consider provides
only phase values, and must be used in conjunction with one of the other methods. 
Detailed descriptions of all these methods are provided in the sections below.

A few general conclusions can be noted now: 
First, it is clear that there is no universal answer for which method is best, since it depends
on what parameters are required and what phase values are of interest. 
Second, offset noise generally
has more impact than phase noise with the same $\sigma$, 
which is visually apparent already in Fig.~\ref{fig_examples}. 
We assume here that no foreknowledge is available about the
type and magnitude of noise affecting a data set, but when such information is available, it can be used
to guide the analysis method. 
Third, when operating at $\phi = \pi/2$ to ensure zero bias, offset noise
yields a phase deviation about three times larger than the projection noise limit (and thus six
times larger than the limit for a standard interferometer with the same number of atoms). In contrast,
for phase noise it is possible to reach within 25\% of this limit. 

Note that the sensitivity to bias is limited by our numerical resolution. If a method gives
phase deviation $\delta\phi$, then with $K$ samples we cannot discern a bias smaller than
$\delta\phi/\sqrt{K}$ due to statistical noise. This can be seen in the statistical values (blue points)
in Figs.~\ref{fig_phase}(b) and (e), where the amplitude deviation is 
$\delta a = (0.1/\sqrt{50}) \delta\tilde a \approx 0.05$, and the amplitude bias results fluctuate by roughly
$0.05/\sqrt{10^4} = 5\times 10^{-4}$.

\begin{table}
\centering
	\begin{tabular}{lcll}
    Method & Relative time & & Scaling \\ \hline
    Statistical & 1 & & constant \\
    Bookstein & 2 & & constant \\
    Geometric & 10 000 & & $n\sigma^{1/2}$ \\
    Reciprocal & 1 000 & & $n$ \\
    Bayesian & 2 000 & & $n^{3/2}/\sigma$    
	\end{tabular}
	\caption{\label{tab_times}
    Relative empirical run times for the analysis methods and parameter values presented in Figs.~\protect\ref{fig_offset}
    and \protect\ref{fig_phase}. 
    For absolute reference, the statistical method requires about 10~\textmu s on a Windows desktop computer
    with a 3.5~GHz i5 CPU.  The `Scaling'
    entry describes how the run time varies with the number of data points $n$ and noise level $\sigma$. The scaling
    indicated is relevant to ranges of roughly $20 \leq n \leq 200$ and $0.01 \leq \sigma \leq 0.1$. In particular,
    the statistical and Bookstein methods scale as $n$ for large $n$, but at the values considered here, fixed
    computational overheads are more significant.  
    }
\end{table}

\section{Fast Methods} \label{secalg}

\subsection{Statistical Method} \label{stats_section}

The statistical and algebraic methods are straightforward linear analysis techniques
that can be performed in real time. 
In general, statistical methods are characterized by
the statistical moments of the data set and have been used in image analysis to constrain geometrical parameters \cite{teague1980image}.


For the interferometer signal model in Eq.\eqref{model}, the ellipse parameters can be estimated from the first and second moments of the signals:
\begin{alignat}{3} \label{stats}
& c_x  = \langle x \rangle & \qquad & c_y = \langle y \rangle \nonumber \\
& a_x  = \sqrt{2 \var(x)} & \qquad & a_y = \sqrt{2 \var(x)} \nonumber \\
& \phi  = \arccos \left( \rho_{x,y} \right)  & &    
\end{alignat}
where $\var(x)$ is the variance $\langle x^2 \rangle - \langle x \rangle^2$ and $\rho_{x,y}$ is the correlation coefficient $\rho_{x,y} =$ $ \cov(x,y)/\sqrt{\var(x) \var(y)}$ with covariance
$\cov(x,y) =\langle xy \rangle - \langle x \rangle \langle y \rangle$.
As seen in Figs.~\ref{fig_offset} and \ref{fig_phase}, the statistical method has a phase bias comparable to other methods,
and offers the lowest amplitude bias near the degenerate points $\phi = 0$ and $\pi$ of all considered methods. 
However, compared to other techniques
the statistical method exhibits larger deviations in phase, center and amplitude for a given noise level and number of data points. 
An exception is the performance near the degenerate points, where the precision is on par with that of other methods. Another advantage of the method is its ease of implementation, as most data analysis
platforms provide built-in functions to evaluate the required moments.

In the limits $n\rightarrow \infty$ and $\sigma\rightarrow 0$, it is straightforward to show that the estimators of 
Eq.~\eqref{stats} converge to the exact parameter values from the model equations \eqref{model}. The method can be improved
by expanding the estimators to incorporate the effects of finite $n$ and $\sigma$, as described in Ref.~\cite{Reinhardt2026prep}.
This approach can provide significantly lower bias values, but for best performance it requires relatively large $n$ and 
measurements at multiple values of $\phi$. 

Another approach based on statistical analysis is to construct histograms of the $x$ and $y$ data values and fit them
to the appropriate marginal density to obtain center and amplitude values \cite{Geiger2011, pelluet2025atom}. 
Derr {\em et al.} have recently
shown that it is also possible to retrieve differential phase values through similar analysis of the sum
or difference signals $x\pm y$ \cite{derr2026parameter}. This method can be used even when only the sum signal is available, 
and not individual $x$ and $y$ values. Again, however, large numbers of data points are required to obtain
sufficient statistical significance in the marginal density \cite{alegria2024number}, and measurements at different
phase values are needed to interpret the sum and difference signals. 

In the following, we focus instead on methods that are suitable for analyzing individual data sets 
with a relatively modest number of data points, reflecting the potentially slow measurement rate for
atom interferometers. Specifically, we consider data sets ranging from tens to hundreds of measurements,
whereas the improved statistical approaches mentioned above typically require hundreds of measurements or more.

\subsection{Algebraic Methods}

Algebraic methods use a quadratic equation for the ellipse in the form
\begin{equation} \label{algebra}
Ax^2 + By^2 + 2C xy + Dx + Ey + F = 0,
\end{equation}
and choose the coefficients to minimize the cost function
\begin{equation} \label{algebra_min}
\chi^2 = \sum_{i=1}^n (Ax_i^2 + By_i^2 + 2C x_i y_i + Dx_i + Ey_i + F)^2.
\end{equation}
The phase is then obtained as $\phi = \cos^{-1}(\pm C/\sqrt{AB})$.

\begin{figure}
	\includegraphics[width=\textwidth]{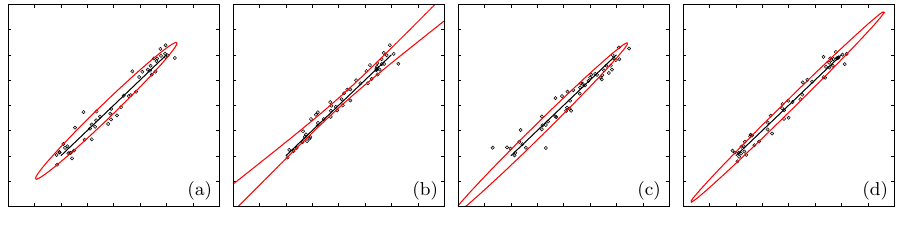}
    \centering
	\caption{\label{alg_examples}
    Example Bookstein fits, for offset noise with $\sigma = 0.1$ and $\phi = 0$. Points are synthetic data, 
    black lines are true (noise free) ellipse, and red curves from the (unmodified) Bookstein method.
    (a) Satisfactory Bookstein fit. 
    (b) Hyperbola returned by the Bookstein fit. (c) Incorrect center returned by the Bookstein fit. (d) Incorrect amplitude
    returned by the Bookstein fit.}
\end{figure}

Because Eq.~\eqref{algebra} is invariant under scaling of the coefficients, \eqref{algebra_min} must be
minimized under some normalization constraint. Different choices for this constraint lead to a broad
family of algebraic methods which have been explored extensively in the literature \cite{Bookstein1979, Gander1994, Zhang1997, fitzgibbon1999direct} .
A popular choice is the Bookstein method \cite{Bookstein1979}, which fixes $A^2 + B^2 + 2C^2 = 1$. The constrained minimization
is carried out by introducing a Lagrange multiplier $\lambda$, and because both \eqref{algebra_min} and the constraint are quadratic
in the coefficients, differentiating them results in a system of linear equations that can be reduced to a generalized eigenvalue 
equation and solved efficiently. Examples of Bookstein analysis are shown in Fig.~\ref{alg_examples},
where part (a) illustrates satisfactory performance.

In contrast, 
plots (b) through (d) show problems with Bookstein method that can occur near the degenerate phases $0$ and $\pi$. As seen in (b),
the fitted conic section is not necessarily an ellipse. 
Even when an ellipse is obtained, the center and/or amplitude can be 
substantially incorrect, as seen in plots (c) and (d). Note that in these cases, the Bookstein fits have not technically failed,
since every data point does lie close to the curve. Rather, the analysis does not account
for the expectation that the data points should be evenly distributed over the entire curve. We suggest a remedy for this
in the context of a geometric method below.

Here, however, we implement a few simple patches to make the Bookstein fits usable in all cases, based on
the observation that the statistical method is not subject to this type of problem and offers reasonably good
performance at degeneracy. The bad Bookstein fits usually occur
when the discriminant $Q = AB-C^2$ is close to zero, so we set an empirical threshold and use the
statistical result instead when $|Q| < 0.002$. After that, we compare the Bookstein center and amplitudes
to the statistical results, and if they differ by more than half the statistical amplitude, we use the statistical result.
Finally, if the previous tests are passed but $Q$ is negative (corresponding to a hyperbola), 
we use the Bookstein center and amplitudes but the statistical phase.

This `modified' Bookstein method was used to generate the data in Figs.~\ref{fig_offset} and \ref{fig_phase}.
We see that method performs comparably to the statistical method in terms of phase bias and 
worse in terms of amplitude bias, but has significantly lower noise in general. Under the conditions
of Fig.~\ref{fig_offset}, at $\phi = 0$ the conventional Bookstein method is rejected 85\% of the time: 
55\% for $Q$ being too small, 1\% for a bad center or amplitude, and 
19\% for returning a hyperbola.
For $\phi > 0.4$~rad, no Bookstein failures were observed under these noise conditions.

An alternative solution to the Bookstein problem is to change the normalization constraint to $AB - C^2 = 1$, 
which ensures that the result is an ellipse. This is known as the Fitzgibbon method \cite{fitzgibbon1999direct}, and while it 
avoids the Bookstein instabilities, it gives a phase bias that is substantially
larger than that of the statistical method so we do not recommend it. The Fitzgibbon method does, however,
give good noise performance for all $\phi$, which could be useful when operating
near the degenerate points. 

Another algebraic technique is the ``self-normalized'' fit of \cite{Harker2006,Collett2014Part1}. This provides a smaller bias
than the Bookstein method, but we find it to fail more often and over a wider range of $\phi$ values. 

\begin{figure}
	\hspace*{-1em}\includegraphics[width=0.5\textwidth]{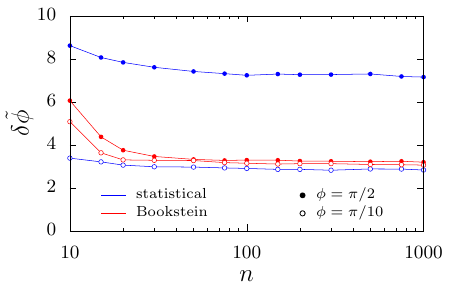}
    \centering
	\caption{\label{algn}
    Dependence of the scaled phase deviation $\delta\tilde\phi$ on the number of data points $n$. All data sets use
    offset noise with $\sigma = 0.1$ and $K = 10^4$. Blue points are obtained with
    the statistical method and red points with the Bookstein method. Closed points are at phase $\phi = \pi/2$ and
    open points at $\phi = \pi/10$.
    }
\end{figure}

Figure~\ref{algn} shows how the statistical and (modified) Bookstein methods perform as a function
of the number of data points. For sufficiently large $n$, the phase deviation uniformly decreases as $1/\sqrt{n}$,
such that the scaled $\delta\tilde\phi$ values are approximately constant. For $n$ below 20 or so, the 
scaled deviations do tend to increase. Similar behavior is observed for
the center and amplitude deviations. In contrast, the phase and amplitude bias are constant at large $n$.

\section{Geometrical Methods} \label{sec:geom}

\subsection{Exact Evaluation}

Geometric methods for ellipse fitting are based on the standard technique of minimizing the
squared distances from the data points to the fit curve. Here we define a cost function
\begin{equation}
\chi^2 = \sum_{i=1}^n d_i^2,
\end{equation}
where $d_i$ is the distance from data point $i$ to the ellipse. This has the advantage over  
algebraic methods of providing a clear physical meaning for the cost function. For instance,
the value of $\chi^2$ is related to the average distance from the data to the curve, and 
can thereby identify situations where
the data might not be consistent with the model being used.

The difficulty of the geometric method is in the computation of $d_i$, which requires solving a quartic
polynomial equation. In practice, it is faster to find this solution numerically than to use the general analytic solution by Ferrari.
Nonetheless, the need to find a solution for each point at each minimization step makes the geometric
method very slow. This is typically inconvenient, but may not be prohibitive 
for an interferometer with a low data rate.

However, as Figs.~\ref{fig_offset} and \ref{fig_phase} show, the performance benefits obtained by the geometric
method are not dramatic. The phase bias is generally lower than in the statistical and algebraic methods,
but the phase noise is comparable to the Bookstein method and, most notably, the performance for the center
and amplitude parameters becomes extremely poor at the degenerate phase values. This reflects the same problem 
as the Bookstein fits in Figs.~\ref{alg_examples}(c) and (d): the cost function is minimized when 
each data point is close to the curve, but there is no penalty for having parts of the curve far
away from all data points. 

\subsection{Discretized Evaluation}

Evaluation of the geometric distance can be sped up through various approximations \cite{Collett2014Part1}, 
but these usually require the noise deviation $\sigma$ to be very small compared to the amplitudes $a_x$, 
$a_y$, which may not be sufficiently satisfied for atom interferometers. 
We present here an alternative method to address both the speed and accuracy problems, by discretizing
the fit curve. We introduce $m$ `discretization points' $\vec{R}_j$ obtained from Eq.~\eqref{model} by setting
$\xi = 2\pi(j-1)/m$ for $j = 1$ to $m$. Then, rather than calculating the distance $d_i$ exactly,
we approximate it as the distance from $\vec{r}_i$ to the nearest
discretization point $\vec{R}_j$. This can be determined by an exhaustive search of $|\vec{r}_i - \vec{R}_j|^2$,
which requires $m\cdot n$ evaluations to determine $\chi^2$. 

\begin{figure}
	\includegraphics[width=\textwidth]{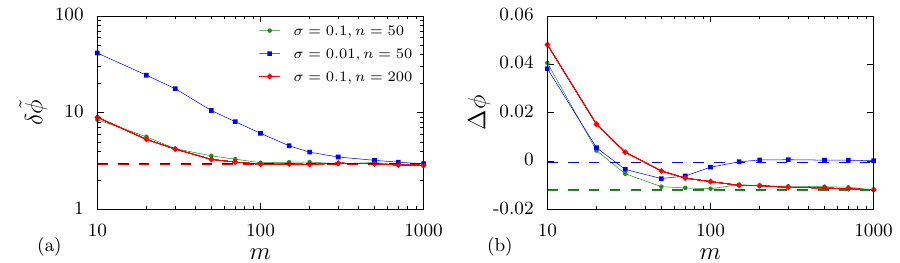}
    \centering
	\caption{\label{varym}
    Dependence of the discretized geometric fit on the number of discretization points $m$. (a) Scaled phase deviation $\delta\tilde\phi$, for offset noise at $\phi = \pi/2$. 
    (b) Phase bias $\Delta\phi$ for offset noise at $\phi = \pi/10$.
    In both plots, point styles and colors correspond to the indicated $\sigma$ and $n$ values, and
    the dashed lines show the corresponding results for continuous geometric fits. All results use
    $K = 2000$.
    }
\end{figure}

It is necessary to choose $m$ appropriately,
since if $m$ is too small the approximation for $d_i$ will be inadequate, while if $m$ is too large
the evaluation will be slow. Figure~\ref{varym} shows how the fit performance depends on $m$, 
with $\delta\tilde\phi$ consistently approaching the value obtained in the continuous calculation.
The $m$ value needed to reach this limit depends strongly on $\sigma$, weakly on $\phi$, and very little on $n$.
We find that an adequate value in all cases is $m = 5/\sigma$,
which can be interpreted
as requiring each data point to have a discretization point within a distance of roughly $\sigma$. 
When using phase noise, the behavior is similar, but the required $m$ is larger, about $10/\sigma$. 
This is consistent with our interpretation: since phase noise causes less geometric spread in 
the data, a larger $m$ value is required to ensure that a discretization point falls within
each data point's noise spread.

In contrast to the deviation $\delta\tilde\phi$, the bias $\Delta\phi$ convergence is more complicated,
as seen in Fig.~\ref{varym}(b). However, once $m$ is large enough for the deviation to converge, the bias variations
are generally small relative to the asymptotic value. It is therefore sufficient to set $m$ based on the deviation and accept that the bias will be somewhat different from
the ideal geometrical result.

\begin{figure}
	\hspace*{-1em}\includegraphics[width=0.5\textwidth]{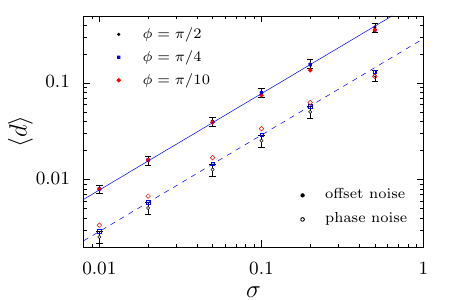}
    \centering
	\caption{\label{sigmafig}
    Estimation of noise level $\sigma$ using mean distance $\langle d \rangle$ from Bookstein fit curve. Each point
    shows the average distance value obtained using $K = 5000$ data sets of $n=50$ points. 
    Solid points use offset noise and open points use phase noise. Point colors and shapes 
    correspond to $\phi$ values as indicated. 
    Representative error bars on the black points show the standard deviations of the $\langle d \rangle$ values.
    Lines show fits of the $\phi = \pi/4$ values at $\sigma \leq 0.2$ to the linear form $\langle d \rangle = \sigma/\kappa$,
    with $\kappa = 1.3$ for offset noise (solid line) and $\kappa = 3.5$ (dashed line).
    }
\end{figure}

From these considerations, it is evident that an estimate for $\sigma$ is needed to determine $m$. 
Previous studies have 
demonstrated that this can be obtained from histograms of the $x$ and $y$ data values,
as discussed in Section~\ref{stats_section}. 
As noted there, however, this is ineffective for small data sets, and it is also 
unable to detect phase noise. We present a more general alternative here:
we first analyze the data set using the (modified) Bookstein method to determine an initial guess fit.
We then use the exact geometric distance calculation to find the distances $d_i$ of each data point from
the fit curve, which is reasonably fast since only one set of distances is required. Finally, we estimate
the noise variance as
\begin{equation} \label{sigmaest}
\sigma\sub{est} = \kappa \langle d \rangle, 
\end{equation}
where $\kappa$ is a calibration factor and $\langle d \rangle$ is the average over the data set. 
For offset noise, we numerically determine $\kappa \approx 1.3$, while phase noise gives $\kappa \approx 3.5$.
These parameters work reasonably well over a wide range of parameters, as illustrated
in Fig.~\ref{sigmafig}. 

In principle, it would be possible to distinguish offset noise from phase noise by combining the histogram method
with our geometrical approach. For our purpose, however, it is convenient that if we assume offset noise
and use the resulting $\sigma\sub{est}$ to calculate $m$, then we will underestimate $\sigma$ for phase noise by
a similar factor as that needed to compensate for the different threshold $m$ values. This reflects
the fact that the required $m$ is fundamentally determined by $\langle d \rangle$ itself. 
We therefore use in all cases
\begin{equation} \label{mlimit}
m = \max\left(\frac{4}{\langle d \rangle}, 20\right),
\end{equation}
where $4 \approx 5/1.3$ and the minimum value of $20$ ensures that the method remains robust even for noisy data.
Using this approach, we obtain performance comparable to that of the exact
geometric method, but running significantly faster except when $\sigma$ is very small.

\subsection{Reciprocal Method}

Besides improving the execution speed, the discrete geometric approach also supports a useful modification
to the cost function. In addition to minimizing the estimated distance from each data point to the fit curve,
we can now also minimize the distance from each discretization point to the data, which prevents the fit
curve from extending into regions far from the data as in Figs.~\ref{alg_examples}(c) and (d). 
(This approach is suitable only when it is known that the variations in the common-mode phase $\xi$
are larger than $2\pi$ so that the data are distributed over the entire ellipse.) We then minimize the cost
function
\begin{equation}
\chi^2 = \frac{1-\lambda}{n}\sum_{i=1}^n d_i^2 + \frac{\lambda}{m}\sum_{j=1}^m D_j^2
\end{equation}
where $D_j$ is the distance from discretization point $\vec{R}_j$ to the nearest data point $\vec{r}_i$,
and $\lambda$ is a weighting parameter such that $\lambda = 0$ corresponds to the ordinary geometrical fit.
For $0< \lambda < 1$, we call this method a reciprocal fit, since it simultaneously fits the data to the ellipse and the 
ellipse to the data. 

\begin{figure}
	\includegraphics[width=\textwidth]{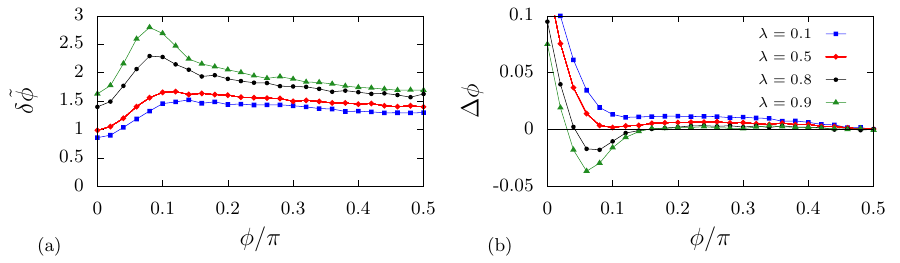}
    \centering
	\caption{\label{varylambda}
    Effect of weighting parameter $\lambda$ on the discrete reciprocal fit. The color and symbol corresond to different $\lambda$ values, as indicated. (a) Scaled phase deviation $\delta\tilde\phi$.
    (b) Phase bias $\Delta\phi$. All points use $n = 50$ and phase
    noise with $\sigma = 0.2$. The black circles ($\lambda = 0.8$) correspond to the value given by Eq.~\protect\eqref{lambda}. 
    }
\end{figure}

We explored various values for $\lambda$, which affect the characteristics of the fit. Figure~\ref{varylambda}
shows some examples. In general, we find that for noisier data, the phase bias can be suppressed 
by using a larger value of $\lambda$,
at the cost of increased deviation. For low-noise data, even a small contribution
from the reciprocal term is sufficient to improve performance. These features are captured heuristically
by the scaling
\begin{equation} \label{lambda}
\lambda = \frac{\sigma\sub{est}^2}{\sigma\sub{est}^2+s_0^2}
\end{equation}
with $s_0 = 0.1$; this is illustrated by the black points in Fig.~\ref{varylambda}. 
We also considered varying $\lambda$ based on the Bookstein estimate for $\phi$.
This can be advantageous if the character of the noise (offset or phase) is known, but for the general case we find \eqref{lambda} to be sufficient. 

The results using this weighting are shown as the violet points in Figs.~\ref{fig_offset} and \ref{fig_phase}.
We see that the reciprocal method generally improves the performance over a pure geometrical fit, and dramatically so for the amplitude
and offset. It retains the advantage of the geometrical fit that the value of $\chi^2$ is related in a clear way to 
the quality of the fit, although it can no longer be interpreted mathematically in terms of the standard chi-squared distribution.

\begin{figure}
	\includegraphics[width=\textwidth]{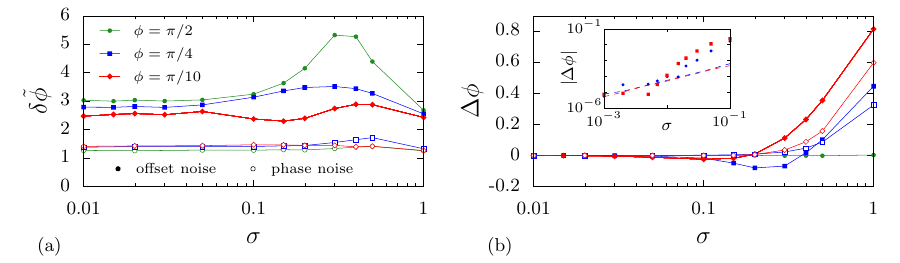}
    \centering
	\caption{\label{varysigma}
    Dependence of the reciprocal fit results on the noise level $\sigma$. (a) Scaled phase deviation $\delta\tilde\phi$ 
    for offset noise (closed points) and phase noise (open points). The color and symbol correspond to
    different $\phi$ values, as indicated. 
    (b) Phase bias $\Delta\phi$ for the same conditions as in (a). The inset shows how the magnitude of the offset-noise bias behaves
    for small $\sigma$. The dashed lines show the corresponding statistical limit of the simulation, $\delta\phi/\sqrt{K}$.
    All plots use $K = 5000$ and $n = 50$. 
    }
\end{figure}

Figure~\ref{varysigma} shows how the phase parameters obtained with the reciprocal method vary with $\sigma$ for
a few representative conditions.
Some structure can be observed in $\delta\tilde\phi$ for large $\sigma$, but the curves typically level out below $\sigma = 0.1$. 
The curves in (b) illustrate that the behavior of the bias can be non-monotonic, and that offset and phase noise
have different character. The inset shows how the magnitude of the bias 
behaves for small $\sigma$. In the range $0.01 < \sigma < 0.1$, the biases
decrease very quickly with $\sigma$. Below 0.01 they become limited to $\delta\phi/\sqrt{K}$ 
by the statistics of the simulation, as shown by the dashed lines.
For phase noise, the bias values are limited by statistics throughout the range of the inset plot.

\section{Bayesian methods}

\subsection{Technique}

A final analysis method we consider is based on Bayes' theorem \cite{von2011bayesian}.
Here we suppose that after some measurements
have occurred, a probability distribution for the phase $P(\phi)$ has been established. If we now make a new measurement, 
obtaining signals $\vec{r} = (x, y)$, the phase probability changes to $P(\phi|\vec{r})$. According to Bayes' theorem, this satisfies
\begin{equation} \label{bayes}
P(\phi|\vec{r})  \propto P(\phi) P(\vec{r} |\phi),
\end{equation}
where $P(\vec{r} |\phi)$ is the probability to measure $\vec{r}$ if we assume that the phase is $\phi$. Given a model
for the interferometer and noise, it is possible to calculate $P(\vec{r} |\phi)$, allowing the determination
of the updated probability distribution. To analyze a data set of $n$ points, we assume initially that
$P(\phi)$ is constant, and then use \eqref{bayes} to successively update the distribution for each point.
At the end, we take the phase estimate to be the value at which $P(\phi)$ is maximum.

This method offers several advantages. If the data and noise models are correct, it provides a mathematically-grounded
way to extract as much information from the measurements as possible. Having access to
the probability distribution $P(\phi)$ makes it easy to evaluate a confidence interval for the estimation.
Finally, it allows the phase estimation to
be updated point by point as new data is acquired, which can be more efficient than reanalyzing the entire data set
as required by other methods (effectively setting $n$ to one in Table~\ref{tab_times}). 

The challenge in the approach is the need for an accurate model. In principle, the offsets, amplitudes, 
and various noise parameters could all be taken as unknowns with their own probability distributions, which could themselves
be updated via the same Bayesian procedure. Unfortunately, the evaluation of $P(\vec{r} |\phi)$ would then
require integration over all these distributions, making it a high-dimensional integral that would be impractical to evaluate. To avoid this problem,
the various parameters can instead be taken as known values. In some situations, it may be possible
to conduct preliminary measurements to determine these values, but such measurements are not trivial and will
impart errors of their own. 

Here we consider a self-contained implementation that obtains the required 
parameters from the data set itself.
To obtain values for the offset and amplitude, we first fit the data using one of the other methods previously
described, and we estimate the noise variance using our previous result Eq.~\eqref{sigmaest}.
To simplify the analysis, we center the data at zero and scale the amplitudes to one. 
In order to calculate the conditional probability distribution $P(\vec{r} |\phi)$, we must also specify a noise model;
we consider either offset or phase noise. Following Stockton {\em et al.} \cite{stockton2007bayesian}, 
the marginal probability distribution for phase noise is
\begin{equation} \label{Bayes_phase}
P(\vec{r} |\phi) \propto \exp\left[-\frac{(\phi-\phi_+)^2}{2\sigma^2}\right] + \exp\left[-\frac{(\phi-\phi_-)^2}{2\sigma^2}\right]
\end{equation}
with 
$
\phi_\pm(\vec{r}) = xy\pm \sqrt{(1-x^2)(1-y^2)}.
$
The phase noise model is straightforward to implement as no numerical quadrature is necessary.
In contrast, in case of offset noise the conditional distribution is obtained through integration \cite{stockton2007bayesian},
\begin{equation} \label{Bayes_offset}
P(\vec{r} |\phi) \propto \int_{-\pi/2}^{\pi/2} G_x(\theta) \big[G_{y+}(\theta)+G_{y-}(\theta)\big]\,d\theta
\end{equation}
where
$
G_x(\theta) = \exp[-(x-\sin\theta)^2/2\sigma^2]
$
and
$
G_{y\pm}(\theta) = \exp\{-[y-\sin(\theta\pm\phi)]^2/2\sigma^2\}
$
are auxiliary functions.
This makes the offset noise model computationally more demanding.
Note that the proportionalities in Eqs.~\eqref{bayes}, \eqref{Bayes_phase} and \eqref{Bayes_offset} 
are sufficient because we can renormalize $P(\phi)$ after each update.\\
For either noise model, we evaluate $P(\phi)$ numerically on a grid. An appropriate grid spacing
can be determined from the phase resolution scaling $\delta\phi \sim \sigma/\sqrt{n}$. To avoid problems
with the grid being too coarse for large $\sigma$ and/or small $n$, we use a spacing of $\Delta\phi\sub{grid} = \pi/(50+\sqrt{n}/\sigma)$.
Further, for the evaluation of the integral in Eq.~\eqref{Bayes_offset}, our grid spacing for $\theta$ is $\Delta\phi\sub{grid}/2$.

\subsection{Performance}

\begin{figure}
	\includegraphics[width=\textwidth]{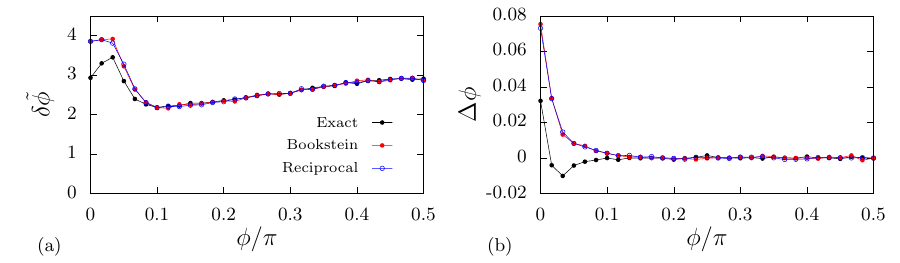}
    \centering
	\caption{\label{bayesparams}
    Effect of parameter estimation on phase deviation $\delta\tilde\phi$ and bias $\Delta\phi$, using the Bayes offset noise model. 
    For all points, $k = 5000$ data sets were constructed using
    offset noise and $\sigma = 0.1$. For the black points, exact values
    for the offsets, amplitudes, and $\sigma$ were provided to the Bayes model. For the blue and red points, the offsets and amplitudes were estimated using the indicated fits, and $\sigma$ was estimated using the method of Eq.~\protect\eqref{sigmaest}
    with $\kappa = 1.3$. 
    }
\end{figure}

Figure \ref{bayesparams} shows the performance of the Bayesian offset model for offset noise, using different methods to determine
the required parameters. The black points are obtained by providing the exact amplitudes, offsets, and noise values
that were used to generate the data. In this case the Bayesian analysis performs very well, with good noise performance
and low bias values. The red and blue points show the performance when the amplitudes and offsets are instead
obtained using the Bookstein and reciprocal fits, while $\sigma$ is estimated
via Eq.~\eqref{sigmaest}. As expected, the performance is
worse than when the exact parameters are used, but it does provide a standalone method for analysis. 
The similarity between using the Bookstein and reciprocal parameters arises because the Bayes calculation is more
sensitive to the value of $\sigma$ than it is to the offset and amplitude values, and $\sigma$ is estimated 
the same way for both cases. In the following,
we use the Bookstein fit for the initial parameters since it is faster.

\begin{figure}
	\includegraphics[width=\textwidth]{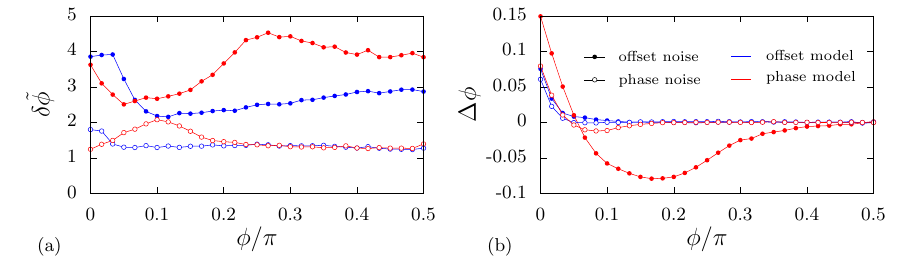}
    \centering
	\caption{\label{bayesmodels}
    Effect of Bayes model type on (a) phase deviation $\delta\tilde\phi$ and (b) bias $\Delta\phi$. 
    For all points, $K = 5000$ data sets were constructed using $\sigma = 0.1$. Filled points used offset noise and
    open points use phase noise. Blue points use the offset model and red points use the phase model.
    }
\end{figure}

Figure \ref{bayesmodels} shows how the phase and offset models perform for both phase and offset noise. For the phase model, we use
$\sigma = 3.5\langle d\rangle$ in the noise estimation, consistent with Fig.~\ref{sigmafig}. While both models handle their intended
noise types well, we find that the offset model (blue) handles phase noise (open points) 
much better than the phase model (red) handles offset noise (filled points).
It is surprising that the offset model performs somewhat better for phase noise than the phase model itself does, but this is
an artifact of the parameter estimation; we do confirm that the phase model handles phase
noise very well when provided with exact parameters. 
However, given the assumed case of no foreknowledge of the parameters or noise type, we find that the
Bayes offset model is superior for general use.

\begin{figure}
	\includegraphics[width=\textwidth]{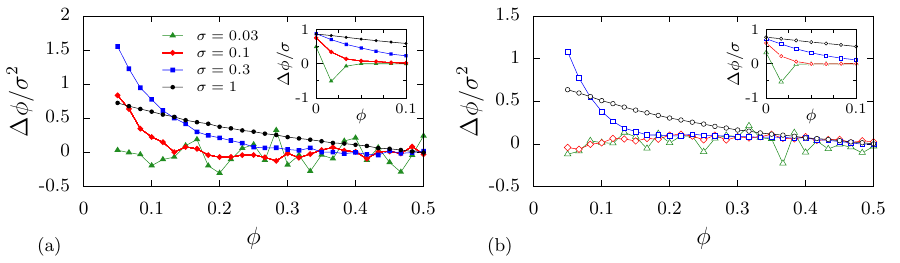}
    \centering
	\caption{\label{bayesperf}
    Behavior of the phase bias $\Delta\phi$ using the Bayesian offset model. All points use $n = 50$ and $K = 5000$.
    Points in (a) are generated with offset noise and in (b) with phase noise. For both plots, 
    the symbol color and shape correspond to different values of $\sigma$, as indicated. 
    The main graphs show larger values of the phase
    $\phi$, where $\Delta\phi$ is scaled by $\sigma^2$ to make the values comparable. In this range,
    for $\sigma \lesssim 0.03$ the bias values are limited by the simulation
    statistics, as seen in the green points. 
    The insets show the behavior for small $\phi$ values, where the bias scales roughly as $\sigma$.
    }
\end{figure}

In fact, Figs.~\ref{fig_offset} and \ref{fig_phase} illustrate that the Bayes offset model performs the best of all the methods considered, except at the degenerate phases where it has good bias performance but relatively large phase deviations. 
Figure \ref{bayesperf} further details how the bias performance varies with $\sigma$. We see that for $\phi \gtrsim \sigma$,
the bias converges quickly to zero as the noise level decreases. For smaller $\phi$, the bias is roughly proportional to $\sigma$,
as expected. As in the other methods, the scaled deviation $\delta\tilde\phi$ does not vary significantly with
$\sigma$ in the ranges considered.

As seen in Table~\ref{tab_times}, the Bayesian method as implemented here is somewhat
slower than the reciprocal model and the run time scales less favorably with $\sigma$ and $n$. The speed
could potentially be improved by using an adaptive grid for $P(\phi)$, as suggested in \cite{stockton2007bayesian}, such that the grid range is
reduced as the probability distribution narrows. 

\section{Conclusions}

The preceding discussions illustrate that the extraction of a differential phase is considerably more
complicated than the extraction of an ordinary phase from a single atom interferometer. 
Each of the methods presented
is appropriate for at least some use cases: The statistical method provides reasonable results
with minimal setup and quick evaluation. 
The algebraic Bookstein method performs better than the statistical method and is nearly as fast, so it
is a good choice for accurate analysis when speed is important. Our implementation does, however, fall back on the
statistical technique for near-degenerate phases to avoid generating hyperbolas in this regime.
When speed is not a concern, the reciprocal method that we have proposed offers better performance,
along with a clear geometrical intuition for the fitting procedure and error interpretation. 
Of the methods considered here, the Bayesian offset-noise model is the most accurate for determining the differential phase,
but it requires another method to provide offset and amplitude estimates.

The Bayesian and reciprocal methods both make use of the noise-estimation technique we have presented in Eq.~\eqref{sigmaest},
based on the mean distance of the data points to the best-fit ellipse. 
In comparison to the competing approach of fitting data histograms to marginal densities, our technique requires fewer data points
and is sensitive to phase noise as well as offset and amplitude noise types. We expect this will be useful in its own
right for characterizing noise effects in differential atom interferometers.


It is notable that the various methods mostly differ in the size of the phase bias $\Delta\phi$,
and how close to the degenerate points it is possible to get before the their sensitivity to phase variations becomes small.
If an experiment is designed to run near the optimal point $\phi = \pi/2$ where $\Delta\phi$ is
generically zero, then the performance of the Bookstein, reciprocal, and Bayesian methods are all
similar. Under offset noise, 
the Bayesian method does have a phase deviation about 15\% lower than that of the Bookstein method,
but whether this is worth the greater run time and computational complexity will depend on the application. 

The discussions here have focused on the question of how to best analyze a differential
interferometer dataset with no extra information available. For high-performance applications, it will
likely be desirable to acquire as much extra information as possible and tailor an analysis
technique to use that information as effectively as possible. While detailing such a procedure
is beyond our scope, the general procedures discussed here can
serve as a useful starting point. For instance, if offset or amplitude parameters can be independently
determined, the values can easily be incorporated into the geometric or Bayesian analyses.

In the context of computer vision, several recent works have explored the use
of machine learning techniques for ellipse detection and characterization
\cite{shi2020robust,long2022ellipse,wang2025efficient}.
In this application, the emphasis is on rapid analysis that is robust to large signal deviations. 
It will be interesting to apply similar techniques to the interferometry context where
errors are typically better controlled but high precision and accuracy are essential. While performance
gains may be available in this way, it will be important to consider the
loss of traceability through the analysis procedure \cite{mora2021traceability}, 
and how that might impact critical results in, for instance, precision measurement applications. 

Finally, we note that the methods considered here are all static, in that they assume the signal
parameters are constant in time. Identifying and processing dynamic parameters, as would be
relevant for applications in inertial navigation, will require additional considerations
\cite{Collett2015Part3,pei2019elementary,Ridley2024}.

Differential atom interferometry is an extremely powerful way to extract useful information 
from a noisy environment, but the signal analysis problem is less straightforward than
might initially be expected. We hope that this paper helps to clarify some of the issues involved and
to introduce some new tools that can be applied to the problem.

%
%

\ack{The authors gratefully acknowledge Jason Boynewicz for early work on the project.}

\funding{This research was funded in part by the National Science Foundation, Grant No.~2110471 and in
part by NASA, Grant No.~80NSSC 24K1237.}


\roles{
\textbf{N. Fowler:}
Conceptualization (equal), Methodology (equal), Software (equal), Validation (equal), Writing – review \& editing (equal).
\textbf{D. B. Reinhardt:}
Conceptualization (equal), Methodology (equal), Software (equal), Validation (equal), Writing – review \& editing (equal).
\textbf{M. Meister:}
Conceptualization (equal), Methodology (equal), Supervision (equal), Writing – review \& editing (equal).
\textbf{C. A. Sackett:} - Conceptualization (equal), Funding acquisition (lead), Methodology (equal), Software (equal), Supervision (equal), Visualization (lead), Writing – original draft (lead).

}












\data{The data used in the figures of this study are available from the corresponding author upon reasonable request.}

\suppdata{Sample Python and Matlab code for the analyses considered here are available in the Supplemental Data.}

\bibliographystyle{vancouver}

\bibliography{biblio}

\end{document}